\documentstyle[12pt,epsfig]{article}
\begin{document}

\title{Function and form in networks of interacting agents}
\author{Tanya Ara\'{u}jo\thanks{%
tanya@iseg.utl.pt} \\
{\small Departamento de Economia, Instituto Superior de Economia e
Gest\~{a}o,}\\
{\small R. Miguel Lupi 20, 1200 Lisboa, Portugal} \and R. Vilela Mendes%
\thanks{%
vilela@alf1.cii.fc.ul.pt} \\
{\small Grupo de F\'{i}sica Matem\'{a}tica, Complexo Interdisciplinar,
Univ.de Lisboa,}\\
{\small Av. Gama Pinto 2, 1699 Lisboa Codex, Portugal}}
\date{}
\maketitle

\begin{abstract}
The main problem we address in this paper is whether function determines
form when a society of agents organizes itself for some purpose or whether
the organizing method is more important than the functionality in
determining the structure of the ensemble.

As an example, we use a neural network that learns the same function by two
different learning methods. For sufficiently large networks, very different
structures may indeed be obtained for the same functionality. Clustering,
characteristic path length and hierarchy are structural differences, which
in turn have implications on the robustness and adaptability of the networks.

In networks, as opposed to simple graphs, the connections between the agents
are not necessarily symmetric and may have positive or negative signs. New
characteristic coefficients are introduced to characterize this richer
connectivity structure.
\end{abstract}


\section{Introduction}

Networks of interacting agents play an important modelling role in fields as
diverse as computer science, biology, ecology, economy and sociology. An
important notion in these networks is the {\it distance} between two agents.
Depending on the circumstances, distance may be measured by the strength of
interaction between the agents, by their spatial distance or by some other
criterium expressing the existence of a link between the agents. Based on
this notion, global parameters have been constructed to characterize the
connectivity structure of the networks. Two of them are the {\it clustering
coefficient} and the {\it characteristic path length}. The clustering
coefficient measures the average probability that two agents, having a
common neighbor, are themselves connected. The characteristic path length is
the average length of the shortest path connecting each pair of agents.
These coefficients are sufficient to distinguish randomly connected networks
from ordered networks and from small world networks. In ordered networks,
the agents being connected as in a crystal lattice, clustering is high and
the characteristic path length is large too. In randomly connected networks,
clustering and path length are low, whereas in small world networks\cite
{Clark} \cite{Watts1} \cite{Watts2} clustering may be high while the path
length is kept at a low level.

An important question is to find out what the mechanisms are, that lead to
each type of structure, when a network of interacting agents evolves in
time. In general, networks of agents organize themselves for some purpose.
For example, a country is organized to insure the survival and well-being of
its inhabitants (or of a subset thereof, anyway), supply networks are
organized to bring food to a town every day and the network of neurons in
the brain is organized to process the information that arrives through the
sensorial organs. Therefore one might be led to think that it is the
function of the network that determines its form. A simple example shows
that it is not necessarily so. Restaurants and private homes in a large city
do not keep more than a few days worth of food and without a continuous
replenishment of their reserves the city would collapse in a few days. The
supply problem of several million inhabitants is solved every day in most
cities by a self-organized network of producers, transporters and retailers
where clustering and a short path length are the rule. Alternatively, in a
centralized economy, a different, very structured system may be organized
with producers delivering their goods to a local cooperative, where they are
collected by a state-organized transportation agency, which then delivers it
to a few centralized stores, where all consumers are supposed to acquire
their goods. In this case one has a very regular structure. People might say
that one system is more efficient than the other, but that is quite
irrelevant as far as functionality is concerned. In both cases the city is
supplied and the fact is, that the structure of the two networks is quite
different.

That institutions used in different societies to achieve similar aims may be
very different is a well known fact. This also casts considerable doubt on
any attempts to characterize the uniqueness of optimal solutions. The
solutions that are arrived at must be largely history-dependent. Any
optimality criterium should therefore not be based on the functionality of
the solution, but on other factors like stability, resistance to change,
adaptability to a changing environment, etc.

Of course, if there is only one possible configuration of the network for
each desired functionality then, whenever the functionality is achieved, the
form is fixed. In that case function determines form and the form does not
dependent on the method by which the functionality is achieved. However,
this is not the most frequent situation in networks of many agents. What we
call the {\it function of the network} is associated to a few collective
variables, like survival of the group, making war on a neighboring country,
maintaining a few simple myths, extracting global concepts like color or
pain from a multitude of external stimuli, etc. That is, the function of the
network is related to a number of variables much smaller than the number of
agents or internal degrees of freedom of the network. In that case it is to
be expected that several distinct configurations of the network will be
associated to the same functionality.

In the space of all the configurations that realize a given function, an
important question is to know what types of network structures do exist,
concerning in particular their connectivity properties (clustering, path
length, etc.). This is the main problem we address in this paper. Because
general statements about networks tend to be vague and do not go a long way,
we concentrate in neural network models that are learning to represent a
given function\footnote{%
Albertini and Sontag\cite{Albertini} have written that on neural networks
''function determines form''. However, they refer to the full detailed
dynamics of the network, not to the input-output binary relationship between
a few nodes that we are considering in this paper.}. The use of neural
networks as a paradigm for networks of interacting agents is not so
restrictive as it may seem because, as shown by Doyne Farmer\cite{Farmer},
they are largely equivalent to many other connectionist systems. The
distance between the agents (nodes) in the network is defined by the inverse
of the absolute value of the connection strength. Nodes are considered to be
connected if the connection strength exceeds a certain threshold. This
threshold is not fixed a priori, but is determined by a clustering algorithm
as the lowest value that insures connectivity of the whole network.

Two learning methods for the same function are tested and, using the
distance defined by the connection strengths, clustering coefficients and
characteristic path lengths are computed. The general conclusion is that, in
fact, function does not determine form, very different structures being
obtained (on average) by the two methods\footnote{%
The $V-$space of the network configurations compatible with a specified
mapping or a given training set, determines what has been called the entropy 
of the network. An adequate control of this quantity is
important for the generalization problem\cite{Denker}. The metric
characterization of the $V-$space given in this paper provides a refiniment
of the residual entropy configurations after the learning process.}. The
first learning algorithm is in the class of reinforcement learning methods,
of which several variants exist\cite{Hebb} \cite{Bak1}, whereas, in the
second, the agents (nodes) are punished by mistakes but nothing happens when
the answer is the desired one\cite{Bak2}. We will denote the first method by 
{\it reinforcement learning method} (RLM) and the second by {\it learning
from mistakes} (LFM).

The neural network, as a network of agents, is richer than the graphs that,
in the past, have been used to study connectivity in networks. This arises
from the fact that not only the connections between nodes may be positive
(excitatory) or negative (inhibitory), but also the connections are in
general asymmetric, one node having an influence over other node different
from the influence it receives from the latter. To characterize this
additional information on the structure of the network we have introduced
new quantities to measure these properties, namely\ {\it symmetry,
cooperation, antagonism and residuality coefficients} as well as {\it %
directed path lengths.}

\section{Clustering and path length in goal-oriented networks}

As a paradigm for goal-oriented networks, we study a neural network which
starts as randomly connected (with small connection strengths) and learns to
reproduce a function by two different learning methods. A certain number of
nodes are defined to be input nodes and some others output nodes. In the
numerical experiments we report here, we have taken two nodes as input, one
as output and the function to be reproduced is a Boolean function like the
exclusive OR, for example. Similar results are obtained for other more
complex functions. The only care to be taken is that the network should have
a sufficiently large number of nodes to guarantee that the subspace of
strength connections, compatible with implementation of that function, is
large. Otherwise, if there is only one possible configuration of connection
strengths, function determines form and the dependence on the learning
method cannot be detected. It is also obvious that, even when the regions in
function space, explored by different learning methods, are distinct, it may
happen that, by chance, configurations obtained by different methods do
coincide. This is borne out in our experiments, some overlap being observed
between the configurations obtained by different learning methods. However,
on average, different methods explore quite different regions of the
function space.

The results we obtain, indicate that the structures resulting from the {\it %
reinforcement learning method} (RLM) exhibit a high clustering coefficient
together with an intermediate value for the characteristic path length. On
the other hand, the structures obtained by the {\it learning from mistakes
method} (LFM) have a low clustering coefficient with characteristic path
lengths similar to those obtained from random structures, that is, similar
to the structures used to initialize the network. RLM seems to be largely
dependent on the establishment of a highly clustered configuration while LFM
does not require the creation of such an ordered structure. One may think of
the structures created by the latter method as being those of a highly
adaptive system where specific tasks are performed without strongly
committed configurations.

\subsection{The network and the learning algorithms}

The network we study is characterized by a non-layered architecture
representing a fully connected system of twelve agents with connection
strengths initially chosen at random in the interval $-0.5<w_{ij}<0.5$.

The absolute value of the connection strength $w_{ij}$ corresponds to the
inverse distance of the agent pair $i$,$j$. From the initial, random,
configuration, two different learning algorithms were used to obtain an XOR
function. The first one is a Hebbian-like method, while the second relies on
the learning from mistakes approach, with reinforcement being replaced by a
process of depressing the synaptic connections involved in mistakes\cite
{Bak2}. Both have a biological inspiration, the first one corresponding to
long term synaptic potentiation\cite{Hebb} and the second to long term
synaptic depression\cite{Crepel}.

In both learning methods, a sigmoid function $\phi $ is used as activation
function and the computation of the output signal includes a bias $\alpha $,
which operates as a regulatory mechanism for the overall activity of the
network. If the network activity is too low, the bias $\alpha $ is decreased
until an appropriate number of firing neurons is obtained. On the other
hand, if the activity exceeds a certain limit, $\alpha $ is increased in
order to keep a low level of activity in the network. A neuron is defined as 
{\it firing} if its output is above $50\%$ of the maximum output ($1$).

{\bf Reinforcement learning method (RLM)}\medskip

The connections between firing neurons are strengthened or weakened
according to whether the output is successful or not. The process affects
all firing neurons in the same way. The reinforcement updating rule is:

If the output is the correct one 
\[
w_{ij}:=w_{ij}+(\delta Y_{i}Y_{j}) 
\]
otherwise 
\[
w_{ij}:=w_{ij}-(\delta Y_{i}Y_{j}) 
\]
$Y_{i}=\phi \left( \sum_{k}w_{ik}Y_{k}\right) $ is the output of neuron $i$
and $\delta <<1$.

Also, as stated before, the bias is adjusted to keep the overall network
activity at a low level. Saturation is avoided by a global rescaling of all
the coupling strengths, when one of them exceeds a fixed threshold.\medskip

{\bf Learning from mistakes (LFM)}\medskip

If the output happens to be the desired one, nothing is done, otherwise the
connections between firing neurons are depressed by a fixed amount $\delta $%
, which is redistributed among the connections between non-firing neurons, 
\[
w_{ij}:=w_{ij}-\delta 
\]
with $\delta <<1$

As in RLM, there is a global regulatory mechanism to keep the overall
activity at a low level.

\subsection{Clustering coefficient and characteristic path length}

Clustering coefficient (CC) and characteristic path length (CPL) are
important statistical parameters used in graph theory. These two parameters
have been the object of growing attention ever since the small-world
phenomenon was identified as an interesting property of the structures found
in many different fields. The small-world feature\cite{Watts2} is
characteristic of structures with CC similar to the one obtained in regular
structures but with CPL's close to those of random networks.

In the past, graph modelling concerned itself mostly with completely random
or completely regular structures. Regular graphs combine high CC with large
CPL while, in the opposite case, random graphs exhibit low CC and small CPL.
Starting from a completely regular structure and applying a random rewiring
procedure to interpolate between regular and random networks it has been
found\cite{Watts1} that there is a broad interval of structures over which
CPL is almost as small as in random graphs and yet CC is much greater than
expected in the random case. This rewiring procedure helps to characterize
the structural aspects of a network in the transition from order to disorder.

In the work presented here, the networks move in the opposite direction,
from completely random towards a goal-oriented structure. The basic
intuition is that forcing a randomly weighted network to learn a function by
different learning methods, may lead to different forms of organization even
though the methods are both targeted at reaching the same functional goal.
In particular, we want to find out to what extent the success of a learning
method is dependent on the transition from disorder to order in the network
structure. The connectivity of the (starting) randomly connected networks
provides reference values that help to characterize the lack of order. In
the other extreme, the more regular structures that arise from learning are
evaluated by finding out how much their CC and CPL differ from those that
characterize randomness in the starting configurations.

CC and the CPL usually apply to graph structures that are connected and
sparse. Since the networks we work with are fully-connected structures, a
first step is targeted at obtaining a sparse representation of the network,
with the {\it degree of sparseness }generated by the learning method itself,
instead of an {\it a priori} specification. Notice that, when looking for a
suitable degree of sparseness, one must avoid disconnected graphs (where the
value of the CPL of any disconnected element would be infinite). For this
purpose we construct of a graph structure from the connection
strengths.\medskip

{\bf The graph representation of the network\medskip }

From the $n\times n$ matrix $W$ of connection strengths, $\left\{
w_{ij}\right\} $, we construct a $n\times n$ distance matrix $D_{W}$, with
elements $d_{ij}=\mid 1/w_{ij}\mid $. Based on the distances $d_{ij}\in
D_{W} $, a hierarchical clustering is then performed using the {\it nearest
neighbor method. }Initially $n$ clusters corresponding to the $n$ agents are
considered. Then, at each step, two clusters $c_{i}$ and $c_{j}$ are clumped
into a single cluster if 
\[
d_{c_{i}c_{j}}=\min \left\{ d_{c_{i}c_{j}}\right\} 
\]
with the distance between clusters being defined by 
\[
d_{c_{i}c_{j}}=\min \left\{ d_{pq}\right\} 
\]
with $p\in c_{i}$ and $q\in c_{j}$

This process is continued until there is a single cluster. This clustering
process is also known as the {\it single link method}, being the method by
which one obtains the minimal spanning tree (MST) of a graph. In a connected
graph, the MST is a tree of $n-1$ edges that minimizes the sum of the edge
distances.

In a network with $n$ agents, the hierarchical clustering process takes $n-1$
steps to be completed, and uses, at each step, a particular distance $%
d_{ij}\in D_{W}$ to clump two clusters into a single one. Let $%
C_{W}=\{d_{q}\}$, $q=1,...,n-1$, be the set of distances $d_{ij}\in D_{W}$
used at each step of the clustering, and $L_{W}=\max \{d_{q}\}$. It follows
that $L_{W}=d_{n-1}$.

At this point we are able to define a representation of $D_{W}$ with
sparseness replacing fully-connectivity in a suitable way. For this purpose,
a boolean graph $B_{W}$ (with $n$ vertices being the network nodes) is
defined setting $b_{ij}=1$ if $d_{ij}\le L_{W}$ and $b_{ij}=0$ if $%
d_{ij}>L_{W}$. As usual, {\it null arcs} of $B_{W}$ are those for which $%
b_{ij}=0$ while for {\it unit arcs} $b_{ij}=1$. Here we want to consider two
nodes $i$ and $j$ to be connected if either $d_{ij}$ or $d_{ji}\le L_{W}$ .
Therefore we take $b_{ij}=\max \{b_{ij},b_{ji}\}$. Later on (Sect. 3) we
will take into account directional effects.

Let $A_{W}$ be the matrix associated with $B_{W}$. Each element $a_{ij}$ is
the number of edges of $B_{W}$ that join the vertices $i$ and $j$ and, since 
$B_{W}$ is a simple graph, $a_{ij}\in \{0,1\}$.

The degree of $B_{W}$, or its {\it coordination number} $k$, represents the
average number of unit arcs leaving each element of the graph. The
coordination number characterizes the sparseness of the graph and has an
important bearing on its properties\cite{Watts2}. In our approach we obtain
the value of $k$ from the network itself. Therefore we avoid any {\it a
priori} estimation and, by the hierarchical clustering method, we also avoid
disconnectivity.

We are also interested in defining $C_{W}^{*}=\{d_{l}\}$, $l=1,...,m$, as
the set of distances $d_{ij}\in D_{W}$ whose values are less or equal to $%
L_{W}$, and computing $r=m-(n-1)$. Clearly $r\ge 0$ is the number of {\it %
redundant} elements in $C_{W}^{*}$, that is, the number of distances $d_{ij}$
that, although being smaller than $L_{W}$, need not be considered in the
hierarchical clustering process. Later on, we will discuss the relation
between the value of $r$ and the clustering coefficient of the graph.{\bf %
\medskip }

{\bf Clustering coefficient\medskip }

The clustering coefficient (CC) of a graph $G$ (averaged over all vertices $%
v $ of $G$) measures whether two vertices adjacent to another vertex $v$ are
adjacent to each other. When CC=1 one has a group of disconnected but
individually complete subgraphs, while CC=0 implies that no neighbor of any
vertex $v$ is adjacent to any other neighbor of $v$.

At the end of the learning process, we build an adjacency list from the
matrix $A_{W}$ associated with $B_{W}$. It is done by listing all vertices
of $B_{W}$ and, next to each one, its neighboring vertices. From the
adjacency list of $B_{W}$ the clustering coefficient of $B_{W}$ may be found
in two different ways.

\begin{enumerate}
\item  The first method computes the value of the clustering coefficient $%
CC_{v}$ of each vertex $v$ by dividing the number of unit arcs in the
neighborhood of $v$ by the total number of arcs in such a neighborhood,
which is given by $s_{v}(s_{v}-1)/2$, $s_{v}$ being the size of the
adjacency list of vertex $v$. Averaging over all vertices of $B_{W}$ we
obtain a coefficient which we denote by $CA_{B_{W}}$.

\item  In the second method the calculation of the clustering of $B_{W}$
does not consider the vertices $v$ of $G$ that have just one vertex in its
neighborhood. For each pair of unit arcs $(v_{1},v_{2})$ and $(v_{2},v_{3})$
of $B_{W}$ that share a common vertex $v_{2}$ we count one if $(v_{1},v_{3})$
corresponds to a unit arc, otherwise we count nothing. The total sum is then
divided by 
\[
\sum_{v=1}^{n}s_{v}(s_{v}-1)/2 
\]
where $s_{v}$ is the size of the adjacency list of each vertex $v$ of $B_{W}$%
. In this way, vertices with a single vertex in its neighborhood do not
contribute to the value computed by the above expression (since $s_{v}-1=0$%
), being those vertices consequently excluded from the computation of $%
CC_{B_{W}}$.
\end{enumerate}

Notice that, for a typical network, the values of $CC_{B_{W}}$ and $%
CA_{B_{W}}$ tend to be very similar. A significant difference between these
values indicates either that the network has many single-neighbor vertices ($%
CC_{B_{W}}>CA_{B_{W}}$) or that the distribution of the clustering
coefficients for each vertex is very heterogeneous ($%
CC_{B_{W}}<CA_{B_{W}}$). To control these effects
we have, for our simulations, computed both $CA_{B_{W}}$ and $CC_{B_{W}}$.

Above, we have defined $C_{W}^{*}=\left\{ d_{l}\right\} $ $\left(
l=1,...,m\right) $ as the set of distances $d_{ij}\in D_{W}$ with values
less or equal to $L_{W}$ and $r=m-(n-1)$ as the number of {\it redundant}
elements in $C_{W}^{*}$, that is, the number of distances $d_{ij}$ that,
although being smaller than $L_{W}$, are not used in the hierarchical
clustering process.

In a connected graph $r$ provides the cardinality of the cycle basis of $%
B_{W}$, or its {\it cyclomatic number}. Being a cycle basis of a graph
defined by the set of its elementary cycles that taken together yield the
entire graph, itself a cycle. In the next section, when discussing the
simulation results, we notice that, depending on the learning method, cycles
and trees (i.e., connected graphs without cycles) may or not appear in the
resulting network structures. The clustered networks have a high
coordination number while in the opposite case the networks approach a
tree-like structure and, consequently, a low clustering coefficient.{\bf %
\medskip }

{\bf Characteristic path length\medskip }

The characteristic path length (CPL) of a weighted graph is the average
length of the shortest path between any two vertices in the graph. From the $%
n\times n$ matrix $W$ of connection strengths, $\left\{ w_{ij}\right\} $,
and its corresponding distance matrix $D_{W}$, the weighted graph $G_{W}$
(with $n$ vertices corresponding to the network nodes) is defined by 
\[
g_{ij}=\min \{d_{ij},d_{ji}\} 
\]

We compute the characteristic path length of a weighted graph $G_{W}$, by
taking for each pair $(i,j)$ in $G_{W}$, with $i\ne j$, the smallest
distance $spl(i,j)$ between $i$ and $j$. The $n(n-1)/2$ edges $g_{ij}$ are
sequentially taken from a list where the $g_{ij}$ were sorted in ascending
order. In the first step, the smallest $g_{ij}$ in the list provides the
shortest distance between $i$ and $j$. In the next steps, the new edge $%
g_{e_{1}e_{2}}$ provides the shortest $spl(e_{1},e_{2})$ distance between $%
e_{1}$ and $e_{2}$ and may also provide another smallest path length by $%
spl(i,j)+spl(e_{1},e_{2})$ if $e_{1}$ or $e_{2}\in \{i,j\}$, namely: 
\[
\begin{array}{l}
\textnormal{if }i=e_{1}\hspace{0.3cm}spl(e_{2},j)=\min
\{spl(e_{2},j),spl(e_{2},e_{1})+spl(i,j)\} \\ 
\textnormal{if }i=e_{2}\hspace{0.3cm}spl(e_{1},j)=\min
\{spl(e_{1},j),spl(e_{2},e_{1})+spl(i,j)\} \\ 
\textnormal{if }j=e_{1}\hspace{0.3cm}spl(e_{2},i)=\min
\{spl(e_{2},i),spl(e_{2},e_{1})+spl(i,j)\} \\ 
\textnormal{if }j=e_{2}\hspace{0.3cm}spl(e_{1},i)=\min
\{spl(e_{1},i),spl(e_{2},e_{1})+spl(i,j)\}
\end{array}
\]

In the following steps we check whether the edge being considered, composed
with the previously established minimal paths, defines a path $spl^{*}(i,j)$
that is smallest than a previously computed $spl(i,j)$. If it happens $%
spl^{*}(i,j)$ replaces $spl(i,j)$.

This computation is sequentially repeated until the shortest distance $%
spl(i,j)$ between each pair of nodes in the graph is obtained. Averaging
over the $n(n-1)/2$ edges of $G_{W}$, we obtain the CPL of $G_{W}$.\medskip

{\bf Results}\medskip

The results presented here were obtained from several simulations in
networks which start as randomly connected. A typical random network is
shown in Fig. 1. In this figure, the absolute value of each connection
strength ($w_{ij}$) of the network specifies the grey intensity of the
corresponding patch in the image. White patches represent null connections
(null arcs). Units 1 and 2 are taken to be inputs and unit 12 is the output.
This is the reason why the other units do not connect back to 1 and 2 and
unit 12 does not connect back to the others.

The absolute values of the connection strengths corresponding to the inverse
of distances, dark patches represent small distances. With connection
strengths chosen in the interval $-0.5<w_{ij}<0.5$, an almost black patch
means $d_{ij}\sim 0$, while an white patch corresponds to $\mid d_{ij}\mid
\sim $ 0.5.

\begin{figure}[htb]
\begin{center}
\psfig{figure=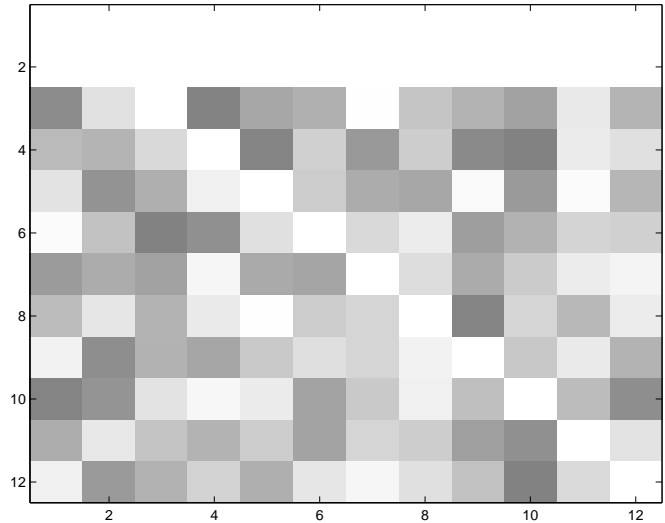,width=9truecm}
\end{center}
\caption{A typical random network}
\end{figure}

The connectivity of random networks provides reference values to
characterize the goal-oriented structures that are obtained by the learning
methods. For this purpose, Fig. 2 shows the image of the adjacency matrix of
a typical random network. It was obtained by:

\begin{enumerate}
\item  Taking the network structure represented in Fig.1

\item  {\it \ }Applying the hierarchical clustering process to obtain the
distance $d_{n-1}\in C_{W}$ used in the last step of hierarchical clustering

\item  Building the corresponding boolean graph with adjacency matrix shown
in Fig. 2. Unit arcs ($d_{ij}\le d_{n-1}$) are represented as black patches
while null arcs ($d_{ij}>d_{n-1}$) correspond to white ones.
\end{enumerate}

\begin{figure}[htb]
\begin{center}
\psfig{figure=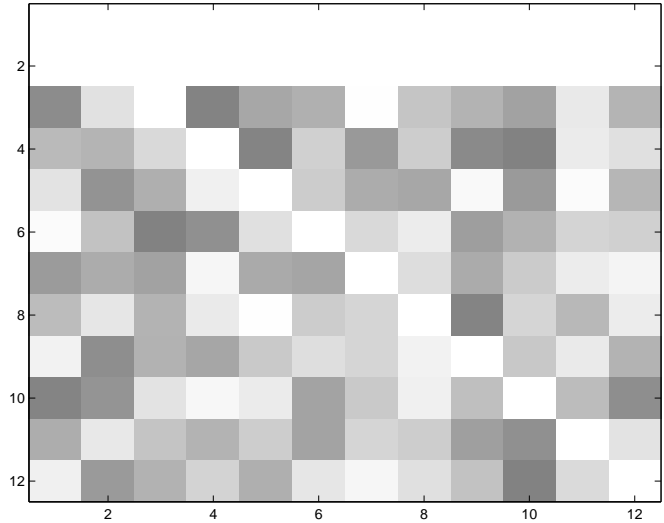,width=9truecm}
\end{center}
\caption{Adjacency matrix of a typical random network}
\end{figure}

As shown in Fig. 2, the graphs that represent the random structures used to
initialize the network are characterized by a high degree of sparseness (a
small number of black patches in the corresponding image). The degree $k$ of
the graph is much smaller than the number of agents in the network.
Moreover, in these graphs the number of unit arcs $u$ is usually only
slightly larger than $n-1$, which is the minimum value that ensures
connectivity. As a consequence, the number of {\it redundant } elements in
the graph is small and the graph approaches a tree-like structure, with a
small clustering coefficient (see table 1).

Fig. 3 shows the typical final structure of a network that starts as
randomly connected and is organized by learning from mistakes (LFM). The
image shows that LFM networks are similar to the typical random structure
shown in Fig. 1. The distribution of connection strengths is not, in
general, very different from those generated at random, suggesting that LFM
does not require the creation of a very organized structure in order to
reach its functional goal.

\begin{figure}[htb]
\begin{center}
\psfig{figure=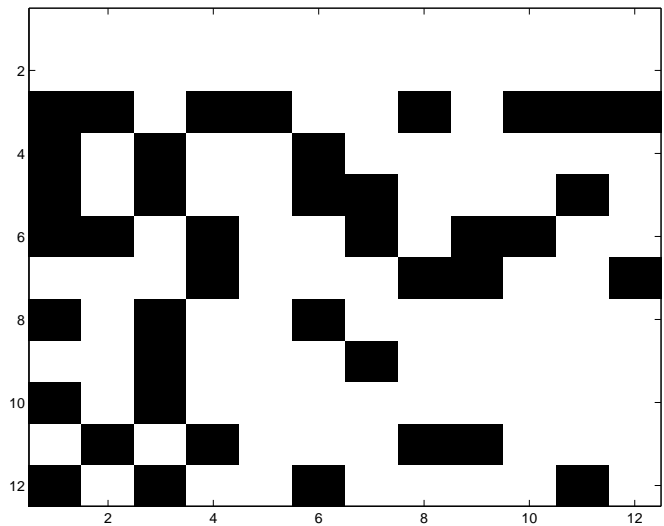,width=9truecm}
\end{center}
\caption{Typical LFM network}
\end{figure}

The image shown in Fig. 4 was obtained following the same steps as used for
the image in Fig. 2. It is built from the network shown in Fig.3. It
represents the adjacency matrix of a typical LFM network. Given that the
typical LFM structures differ little from a random configuration, it
exhibits a significant degree of sparseness. Looking at Fig. 4 we see that
the number of unit arcs ($u$) remains close to $n-1$, hence the number of 
{\it redundant } elements in the graph is almost as small as in random
networks. In a significant number of simulations, the final structures are
even closer to tree-like structures, with a consequently low clustering
coefficient.

\begin{figure}[htb]
\begin{center}
\psfig{figure=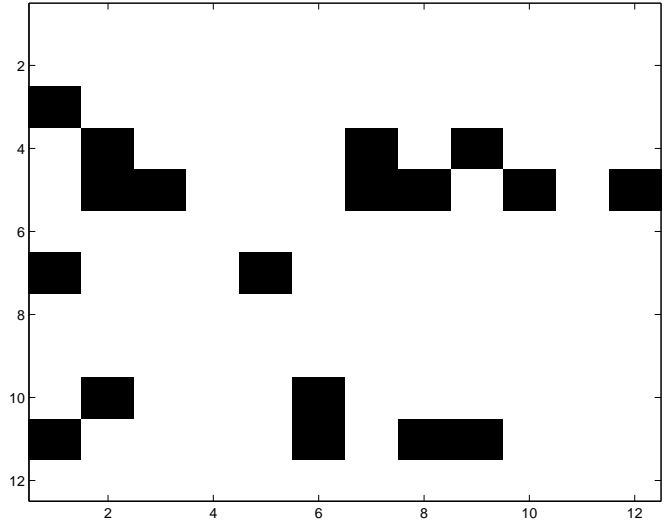,width=9truecm}
\end{center}
\caption{Adjacency matrix of a typical LFM network}
\end{figure}

\begin{figure}[htb]
\begin{center}
\psfig{figure=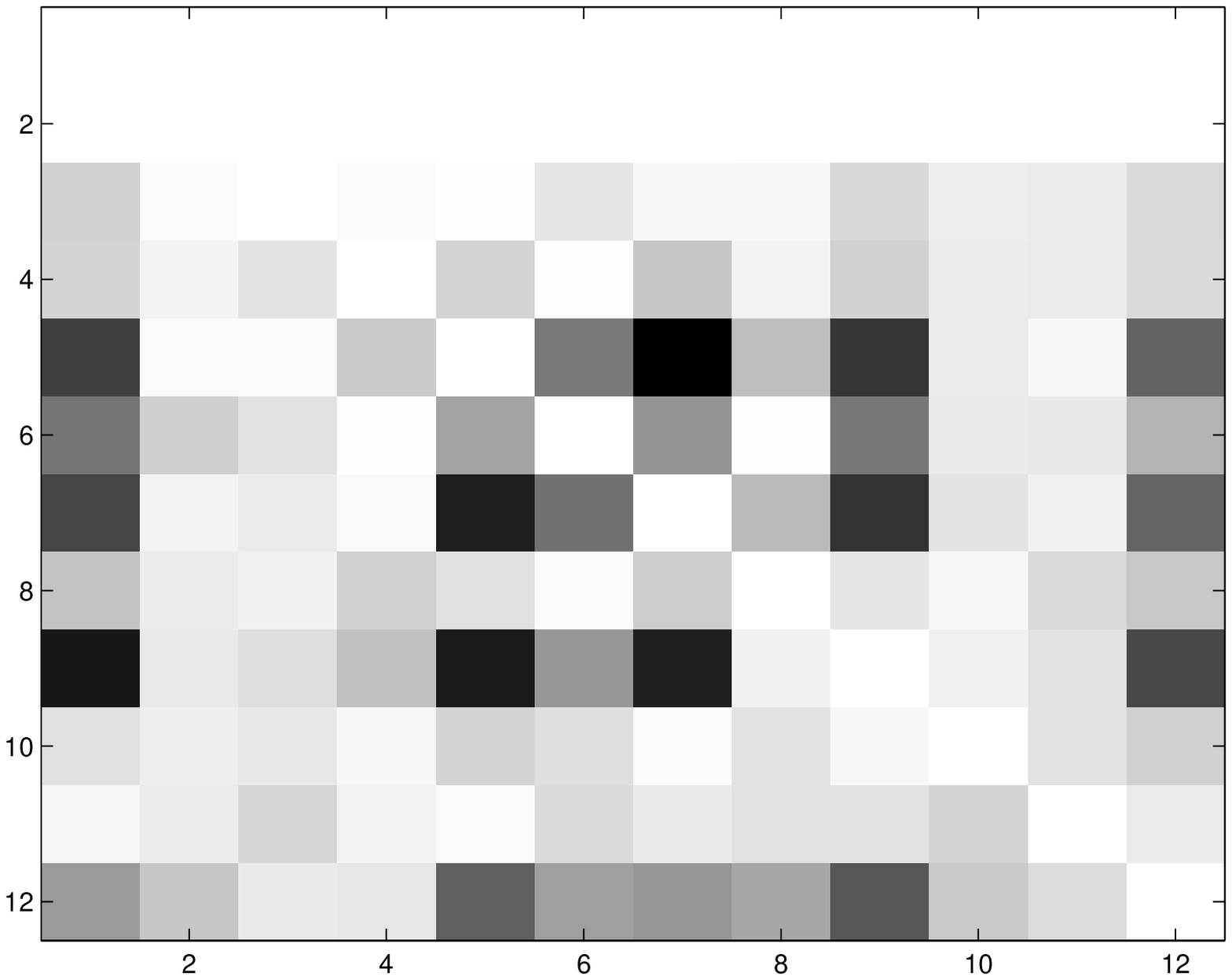,width=9truecm}
\end{center}
\caption{Typical RLM network}
\end{figure}

Fig. 5 shows a typical configuration for a network that learned through RLM.
Small and large distances are not so well distributed as they were at
random, showing that RLM networks move away from the initial configuration
in order to reach its functional goal. The degree of sparseness of Fig. 6
confirms this fact. The degree of sparseness of a typical RLM structure is
smaller than that of a random one and also smaller than the degree of
sparseness of a typical LFM network. Some of the connection intensities are
strongly increased during the learning process and the final network very
often presents a significant degree of symmetry.

\begin{figure}[htb]
\begin{center}
\psfig{figure=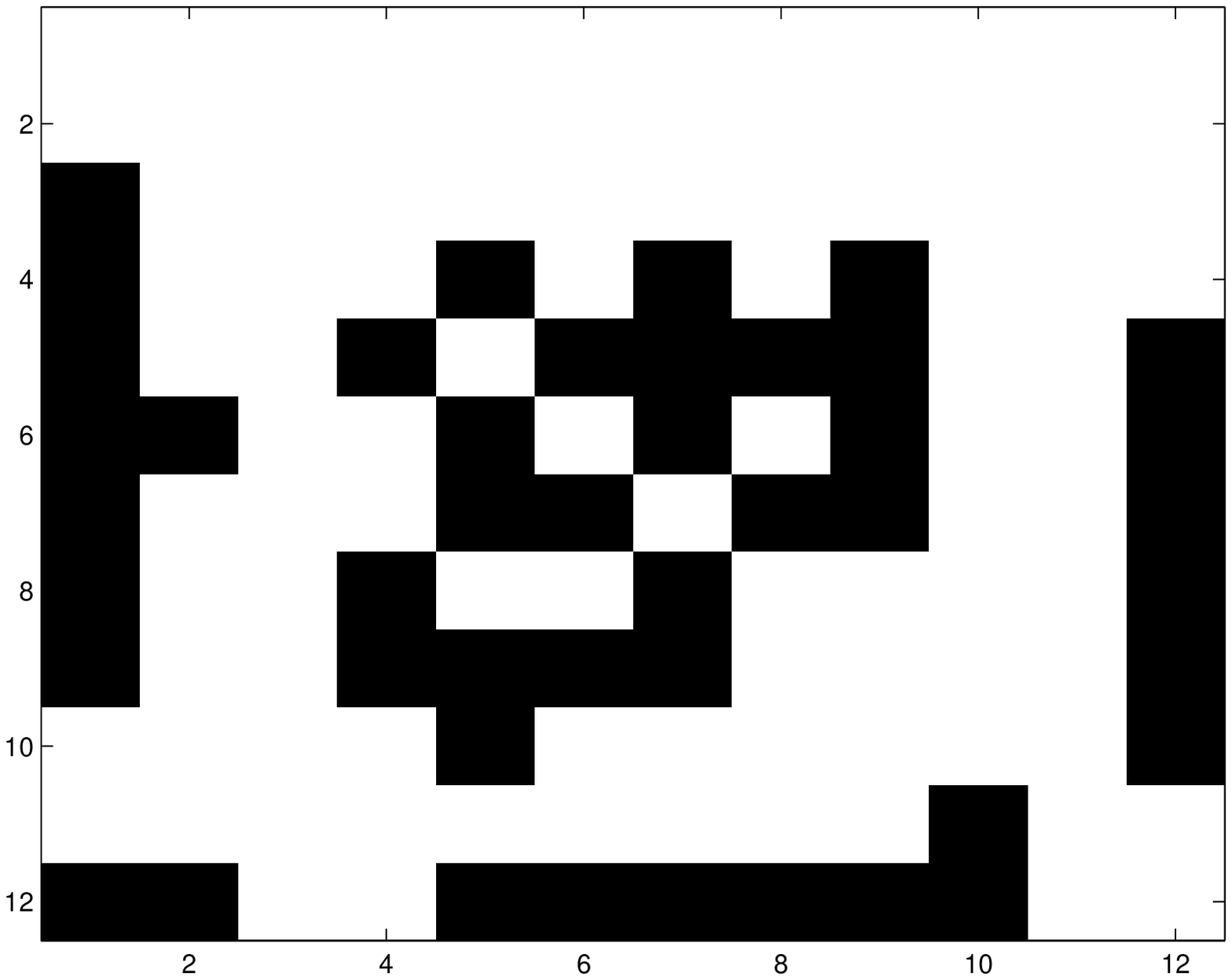,width=9truecm}
\end{center}
\caption{Adjacency matrix of a typical RLM network}
\end{figure}

The number of black patches strongly increases in the structure represented
in Fig. 6 showing that the number of unit arcs ($u$) is much greater than $%
n-1$. Consequently, the number of {\it redundant} elements in the graph is
significantly greater than in random networks. As shown in this figure, the
final RLM structures tend to contain cycles and move away from the tree-like
structures that appear in random and LFM networks.

Table 1 shows typical values for the degree of the graphs ($k$),the
clustering coefficient (CC and CA) and the characteristic path length (CPL)
for random, LFM and RLM networks. In each case we show the mean (\={x}) and
the standard deviation ($\sigma _{x}$) obtained in the simulations.\vspace{%
0.3cm}

\begin{center}
\begin{tabular}{|c|cccccccc}
& $\overline{CC}$ & $\sigma _{CC}$ & $\overline{CA}$ & $\sigma _{CA}$ & $%
\overline{CPL}$ & $\sigma _{CPL}$ & $\bar{k}$ & $\sigma _{k}$ \\ \hline
Rand & 0.26 & 0.16 & 0.24 & 0.17 & 1.6 & 0.05 & 3.3 & 1.0 \\ 
LFM & 0.25 & 0.15 & 0.35 & 0.20 & 2.1 & 0.2 & 3.0 & 0.69 \\ 
RLM & 0.57 & 0.19 & 0.53 & 0.16 & 5.6 & 1.16 & 4.5 & 1.3
\end{tabular}
\vspace{0.15cm}

Table 1: k, CC and CPL typical values\vspace{0.3cm}
\end{center}

Figs. 7 and 8 show the distributions of the clustering coefficient and the
characteristic path length for random, LFM and RLM networks. As mentioned in
the introduction, even though the regions in function space explored by
different learning methods are distinct, it happens that, by chance,
configurations obtained by different methods may coincide.

\begin{figure}[htb]
\begin{center}
\psfig{figure=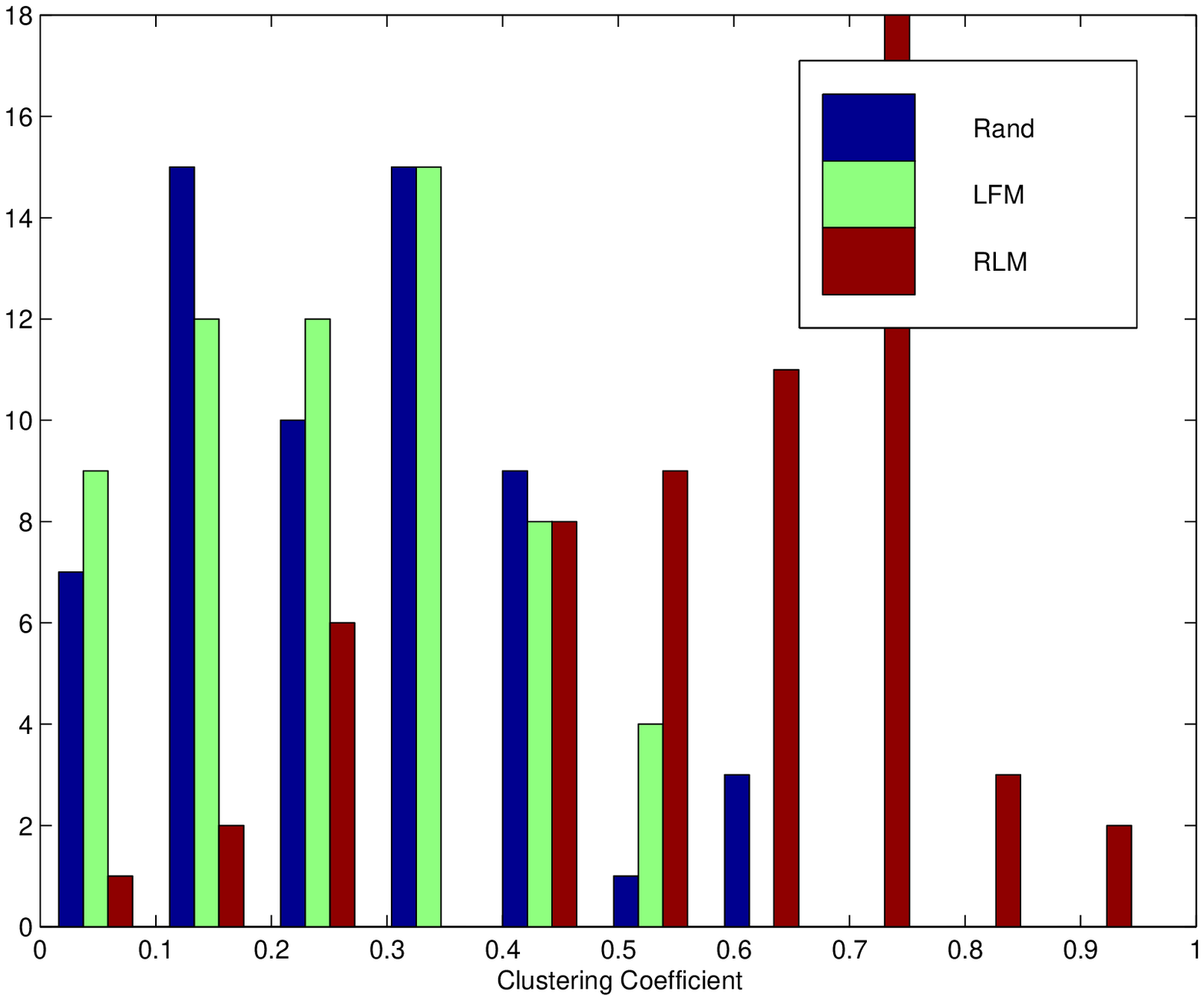,width=9truecm}
\end{center}
\caption{Custering coefficient distribution}
\end{figure}

The histograms of both CPL and CC confirm that there is some overlap between
the configurations obtained by different learning methods. However, on
average, as far as clustering and characteristic path length are concerned,
LFM and RLM exhibit quite different structures.

\begin{figure}[htb]
\begin{center}
\psfig{figure=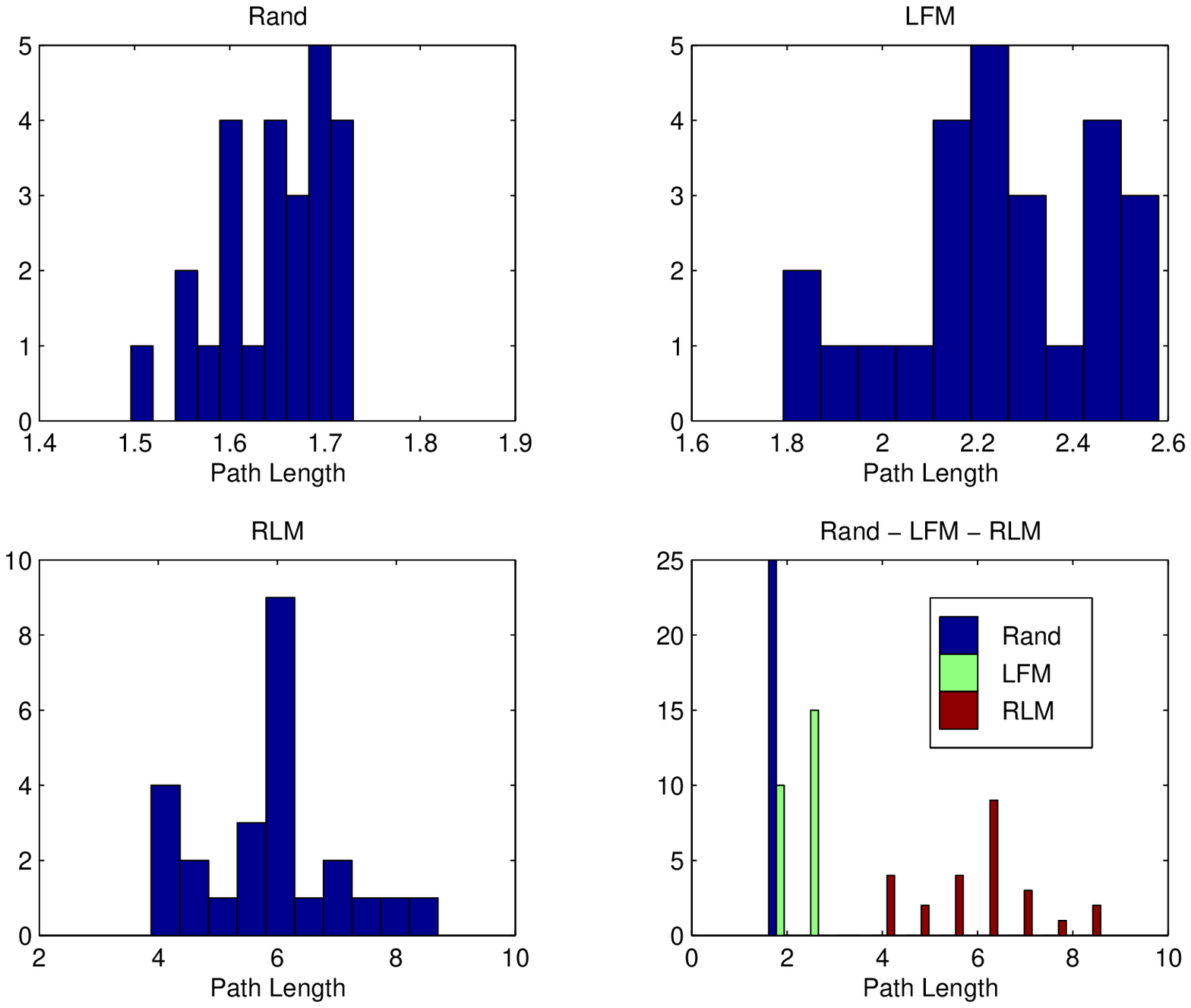,width=9truecm}
\end{center}
\caption{Characteristic path length distribution}
\end{figure}

On the other hand, as far as these parameters are concerned, LFM networks
have structures similar to random networks.

\section{Directional network coefficients}

The coefficients introduced in this section aim at characterizing the richer
connectivity structure of the learning networks because the strengths of
interaction between agents (and their corresponding spatial distances) are
not necessarily symmetric and may have positive or negative signs.

\subsection{Directed path length}

The directed path length (DPL) of a weighted graph provides the average
length of the shortest directed path between any two vertices in the graph.
In order to compute the directed path length of a network we take the $%
n\times n$ matrix $W$ of connection strengths, $\left\{ w_{ij}\right\} $,
and its corresponding distance matrix $D_{W}$, $d_{ij}=\mid 1/w_{ij}\mid $.
The weighted and directed graph $dG_{W}$ is defined by setting

\[
dg_{ij}=d_{ij} 
\]

As in the computation of the characteristic path length, the ($n(n-1)-2$)
edges are sequentially taken from a list with the $dg_{ij}$ sorted in
ascending order. In the first step, the smallest $dg_{ij}$ in the list
provides the shortest distance between $i$ and $j$. In the following steps,
each new edge in the list plays a double role: $dg_{e_{1}e_{2}}$ provides a
distance $dpl(e_{1},e_{2})$ between $e_{1}$ and $e_{2}$ by $%
dpl(e_{1},e_{2})=g_{e_{1}e_{2}}$ and it may also provide a smaller distance $%
dpl(e_{1},j)$ or $dpl(i,e_{2})$ whenever $e_{1}=j$ or $e_{2}=i$. Namely, 
\[
\begin{array}{l}
\textnormal{if }i=e_{2}\hspace{0.3cm}dpl(e_{1},j)=\min
\{dpl(e_{1},j),dpl(e_{1},e_{2})+dpl(i,j)\} \\ 
\textnormal{if }j=e_{1}\hspace{0.3cm}dpl(i,e_{2})=\min
\{dpl(i,e_{2}),dpl(e_{1},e_{2})+dpl(i,j)\}
\end{array}
\]

At each step one checks whether the edge being considered defines a path $%
dpl^{*}(i,j)$ that is smaller than a previously computed $dpl(i,j)$. If it
happens $dpl^{*}(i,j)$ replaces $dpl(i,j)$.

This computation is sequentially repeated until all the $n*(n-1)-2$ edges in
the list are considered. In so doing, the smallest distance $dpl(i,j)$
between all pairs of nodes in the graph is obtained. Averaging over the $%
n(n-1)-2$ edges of $dG_{W}$, we obtain the DPL of $dG_{W}$.\medskip

{\bf Results}\medskip

Figs. 9 shows the distributions of the directed path length for random, LFM
and RLM networks.

\begin{figure}[tbh]
\begin{center}
\psfig{figure=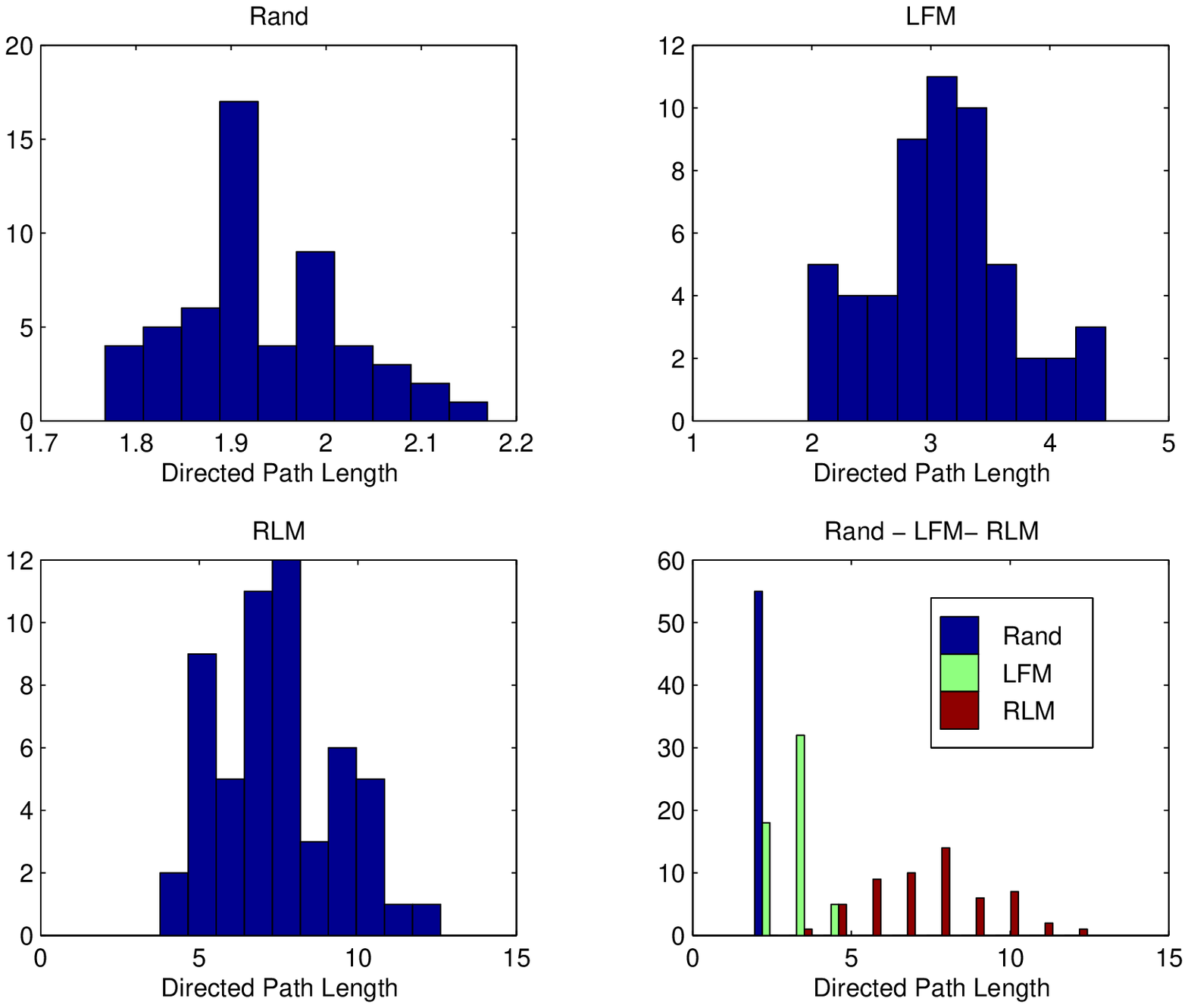,width=9truecm}
\end{center}
\caption{Directed path length distribution}
\end{figure}

Table 2 shows typical values for the degree of the graphs ($k$) and the
directed path length (DPL) for random, LFM and RLM networks. In each case we
show the mean (\={x}) and the standard deviation ($\sigma _{x}$) obtained in
the simulations.

\begin{center}
\begin{tabular}{|c|cccc}
& $\overline{DPL}$ & $\sigma _{DPL}$ & $\bar{k}$ & $\sigma _{k}$ \\ \hline
Rand & 2.1 & 0.7 & 3.3 & 1.0 \\ 
LFM & 3.0 & 2.3 & 3.0 & 0.69 \\ 
RLM & 7.5 & 5.6 & 4.5 & 1.3
\end{tabular}

Table 2: $k$ and DPL typical values
\end{center}

When the orientation of the edges is taken into account, the average length
of the shortest path in random, LFM and RLM networks naturally increases.
The results show that the three types of networks exhibit a similar increase
as compared to the previously obtained CPL values (see table 1). It is still
between random and LFM networks that we find closer values. The typical RLM
networks have a higher DPL, showing that, when the directions of the edges
are considered, a RLM network still exhibits properties that characterize
structures away from randomness. It is in accordance with the idea that the
success of its underlying method is much more dependent on the acquisition
of an ordered structure than the learning by mistakes method.

\subsection{ Symmetry, cooperation, antagonism and residuality coefficients}

As opposed to simple graphs, the connections between the agents in the
networks that we have been studying (and in most natural occurring networks)
are not symmetric and may have positive or negative signs. It is therefore
convenient to be able to characterize this richer connectivity structure.
For this purpose some new coefficients are defined. A {\it symmetry
coefficient} ($S$) is defined by 
\[
S=1-\sum_{i>j}^{n}\mid w(i,j)-w(j,i)\mid /\sum_{i>j}^{n}max(\mid w(i,j)\mid
,\mid w(j,i)\mid ) 
\]
It follows that $-1\le S\le 1$. From the value of $S$ we are able to
evaluate how far the learning networks are from a perfectly symmetric
structure $(S=1)$ and which learning method contributes more to changing the
values $(S\sim 0.5)$ that characterize a typical random network. The results
in table 3 show that on average, after learning, the symmetry coefficient
increases both for RLM and LFM networks. As far as symmetry is concerned,
the two methods behave similarly.

In addition we may also define {\it cooperation} ($C$), and {\it antagonism}
($A$) coefficients by 
\[
C=\sum_{w(i,j)>0}^{n}w(i,j)/\sum_{i\ne j}^{n}\mid w(i,j)\mid 
\]
\[
A=-(\sum_{w(i,j)<0}^{n}w(i,j)/\sum_{i\ne j}^{n}\mid w(i,j)\mid ) 
\]
with $C+A=1$.

Initially, the networks are initialized at random in the interval $%
-0.5<w_{ij}<0.5$. The randomly chosen connections tend to be uniformly
distributed between positive and negative strengths $(C\sim A\sim 0.5)$. One
may think of the positive connection strengths as representing cooperation
between agents, while the negative ones represent antagonistic interactions.
The highest degree of cooperation (and the lowest of antagonism),
corresponding to $C=1$ (and $A=0$), is reached when every network connection
has a positive sign. Conversely the lowest degree of cooperation ($C=0$, $%
A=1 $) is characteristic of a network where every connection strength is
negative.

The last coefficient we will define is the {\it residuality} ($R$)
coefficient

\[
R=\sum_{1/\left| w(i,j)\right| >L_{W}}\mid w(i,j)\mid /\sum_{1/\left|
w(i,j)\right| \le L_{W}}\mid w(i,j)\mid 
\]
where $L_{W}$ is the highest threshold distance value $\left|
1/w(i,j)\right| $ that insures connectivity of the whole network in the
hierarchical clustering process of Sect. 2.1. Residuality relates the
relative strengths of the connections above and below the threshold value.

Table 3 shows the average values obtained for the symmetry (S), cooperation
(C), antagonism (A) and residuality (R) coefficients in random, LFM and RLM
networks.

\begin{center}
\begin{tabular}{|c|cccc}
& $S $ & $C$ & $A $ & $R $ \\ \hline
Rand & 0.57 & 0.51 & 0.49 & 2.6 \\ 
LFM & 0.70 & 0.94 & 0.05 & 1.8 \\ 
RLM & 0.75 & 0.66 & 0.33 & 0.6
\end{tabular}
\vspace{0.15cm}

Table 3: Symmetry, cooperation, antagonism and residuality
\end{center}

The results show that, before learning, in the random networks the weight of
the connections below the threshold $1/L_{W}$ is two to three times higher
than the weight of the connections above the threshold. After learning the
residuality coefficient decreases in both the LFM and RLM networks, with a
very significant decrease in the RLM networks. This is due to the fact that
RLM networks become less sparse after learning (see $\bar{k}$ in table 2)
forcing several {\it residual connections} to leave this category. For the
LFM networks, although sparseness does not change much after learning, the
decrease of $R$ happens because, the connection strengths above $1/L_{W}$
tend to be stronger than those that remain below the threshold.

Cooperation (and antagonism) behaves quite differently depending on the
learning method. In LFM networks, $C$ approaches $1$ after learning, while,
in a typical RLM network, the value of the cooperation coefficient stays
around $2/3$. Antagonism seems to disappear on LFM learning. On the other
hand, RLM learning keeps a reasonable degree of antagonism $(A=0.33)$ in the
network structure.

\section{Robustness and adaptability}

The networks we have studied acquire a structure while learning a function.
While clustering and path length bring information on the connectivity of
the structures, the characterization of the mechanisms leading to each type
of structure raises a few other questions, namely:

(i) How easily will the acquired structure adapt itself to the
representation of another function?

(ii) To what extent do the structures succeed in keeping the same
functionality when some of their connections are suppressed?

As a first step to answer these questions we have measured the {\it %
adaptability} of RLM and LFM networks as follows:

\begin{enumerate}
\item  A network with connection strengths chosen at random in the interval $%
-0.5<w_{ij}<0.5$, learns to reproduce the exclusive OR function

\item  After learning the matrix $x_{ij}$ keeps the resulting normalized $%
\left( x_{ij}=w_{ij}/max(w_{ij})\right) $ connection strengths

\item  The network with connection strengths $x_{ij}$ learns to reproduce
the AND function

\item  After learning we obtain the matrix $a_{ij}$ of the resulting
normalized connection strengths

\item  The network {\it adaptability coefficient} $\gamma _{N}$ is computed
by 
\[
\gamma _{N}=\sum_{i,j=1}^{n}\mid x_{ij}-a_{ij}\mid 
\]
\end{enumerate}

Averaging $\gamma $ over the results of several simulations we have obtained 
$\overline{\gamma }_{LFM}=30.8$, yields an average change $\Delta
x_{ij}=0.25 $ when LFM is the method that is chosen.

Following the same set of steps as above in order to compute the
adaptability of RLM networks turns out to be quite difficult because step 3
frequently fails. In contrasts with LFM structures, adaptation in RLM
networks is almost absent and new learning is efficient only if one starts
from scratch, i.e., from a randomly connected network structure.

These results indicate that the configurations obtained by different
learning methods behave quite differently as far as adaptability is
concerned. The structures created by the LFM method are those of a highly
adaptive system whereas for RLM the structures that are created seem to
become highly specialized for its purpose.

To evaluate the {\it robustness} of the structures the following algorithm
is applied:

(i) A vector $\left\{ x_{i}\right\} $ of $n(n-1)$ components is defined,
corresponding each components to a particular connection in the network. The
vector is initialized with zeros.

(ii) After the learning process, one cuts each one of the connections in
turn and tests whether the learned function is still reproduced. If the test
fails one adds a one to the corresponding component of the vector $\left\{
x_{i}\right\} $.

(iii) The test is repeated for all the connections and for a certain number
of different networks (60 different networks in our simulations)

(iv) The distribution $P(x)$ of the values stored in the vector $\left\{
x_{i}\right\} $ is plotted.

Figs. 10 and 11 show the results corresponding to LFM and RLM networks with
the same number of trials in each case.

\begin{figure}[htb]
\begin{center}
\psfig{figure=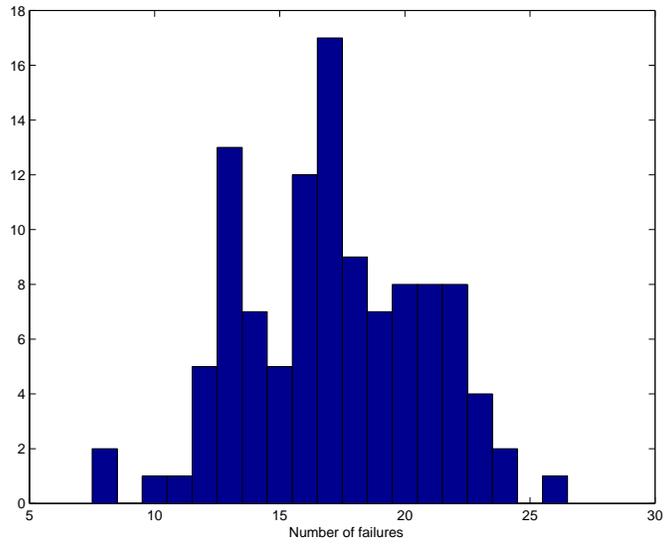,width=9truecm}
\end{center}
\caption{Distribution of failures for LFM networks}
\end{figure}

\begin{figure}[htb]
\begin{center}
\psfig{figure=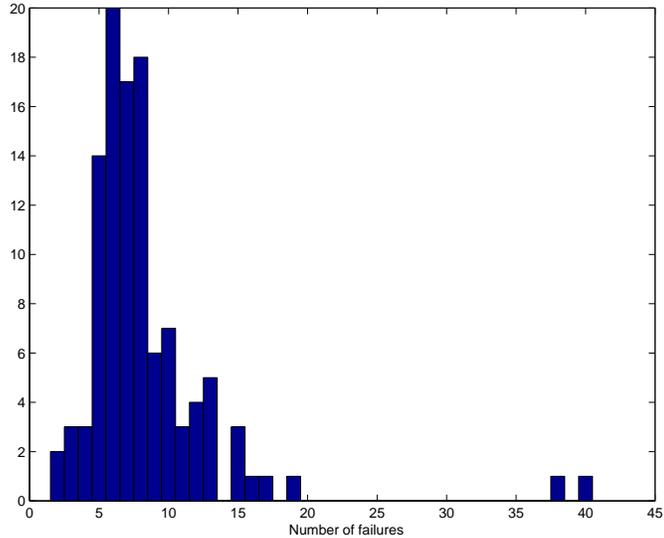,width=9truecm}
\end{center}
\caption{Distribution of failures for RLM networks}
\end{figure}

The results indicate that RLM networks are more robust than those resulting
from the LFM method. The former exhibit, on average, a smaller number of
errors, suggesting that, individually, the role of any specific connection
is less important for reaching the desired functionality than in RLM
networks. Moreover, in the case of RLM networks the distribution of failures
(Fig. 11) shows that some connections are much more important than others
(those that contribute to the fat tail), while a large amount of connections
play a smaller role.

\section{Conclusions}

In multi-agent networks the overall functionality or collective behavior
does not uniquely determine the interaction topology and the graph structure
of the network. This happens because, in general, many different
configurations are associated to the same (small number) of relevant
collective variables. Then, the organizing method, that is, the evolution
history of the network, is the main determining factor on the establishment
of a particular type of structure on the network. These general conclusions
are borne out by our study of networks that, starting from a random
configuration, learn to represent a function by two different learning
methods.

{\it Clustering coefficients} and (non-directed) {\it characteristic path
lengths} turn out to be appropriate to discriminate the two organizing
methods that were used. In particular, a striking confirmation of the ''{\it %
function does not determine form}'' assertion is obtained from the fact that
the high clustering and intermediate path length of RLM networks indicate
that reinforcement learning establishes a highly ordered configuration,
whereas the same functionality is obtained in LFM networks with low
clustering and path lengths similar to random networks.

The idea that learning something or reaching some goal requires some degree
of order is well accepted. So is the knowledge that regular structures - in
opposition to those generated at random - exhibit high clustering and large
path length. Recent work has shown that there is a multitude of cases where
the structures of interest lie in a broad interval between regular and
random. In this paper we have shown that there are cases where the same goal
may be achieved by structures near both extremes. Achieving a goal does not
necessarily require very organized structures. Moreover when the method
followed to achieve the goal implies the establishment of a high degree of
order, the resulting structures tend to be hard to adapt to any different
goal. As shown in the last section of the paper, algorithms may be developed
to characterize, in a quantitive manner, the degree of {\it robustness} and 
{\it adaptability} of the networks.

In the neural-like networks that we have been using (and in most natural
occurring networks) the interactions between the agents are not symmetric
and may have positive or negative signs. This in contrast to the simple
graph structures used in the past to study interaction topologies. {\it %
Directed path lengths}, as well as {\it symmetry}, {\it cooperation}, {\it %
antagonism} and {\it residuality} coefficients were defined, which provide a
refined characterization of the network structures. Relevant differences
were also found between the learning methods when these new coefficients are
measured. For example, starting from a random network, LFM seems to strongly
improve cooperation, whereas in RLM cooperation increases only slightly.

\end{document}